\documentclass{webofc}
\usepackage[varg]{txfonts}   
\usepackage{xcolor, pdflscape}

\def\macs{MACS J0018.5+1626 }
\def\macscom{MACS J0018.5+1626, }

\begin{document}

\title{Improved Constraints on Mergers with SZ, Hydrodynamical simulations, Optical, and X-ray (ICM-SHOX)}

\subtitle{Paper II: Galaxy cluster sample overview}

\author{\lastname{E. M. Silich}\inst{1}\fnsep\thanks{\email{esilich@caltech.edu}} \and
        \lastname{E. Bellomi}\inst{2} \and
        \lastname{J. Sayers}\inst{1} \and
        \lastname{J. ZuHone}\inst{2} \and
        \lastname{U. Chadayammuri}\inst{2} \and
        \lastname{S. Golwala}\inst{1} \and
        \lastname{D. Hughes}\inst{3} \and
        \lastname{A. Montaña}\inst{3} \and
        \lastname{T. Mroczkowski}\inst{4} \and
        \lastname{D. Nagai}\inst{5} \and
        \lastname{D. Sánchez}\inst{6} \and
        \lastname{S. A. Stanford}\inst{7} \and
        \lastname{G. Wilson}\inst{8} \and
        \lastname{M. Zemcov}\inst{9} \and
        \lastname{A. Zitrin}\inst{10} 
}

\institute{Cahill Center for Astronomy and Astrophysics, California Institute of Technology, Pasadena, CA 91125, USA
\and
           Center for Astrophysics $\vert$ Harvard $\&$ Smithsonian, 60 Garden St., Cambridge, MA 02138, USA 
\and
            Instituto Nacional de Astrofísica, Óptica, y Electrónica (INAOE), Aptdo. Postal 51 y 216, 7200, Puebla, Mexico
\and
            European Southern Observatory, Karl-Schwarzschild-Str. 2, D-85748, Garching, Germany
\and
            Physics Department, Yale University, New Haven, CT 06520, USA
\and
            Consejo Nacional de Ciencia y Tecnología-Instituto Nacional de Astrofísica, Óptica, y Electrónica (CONACyT-INAOE), Luis Enrique Erro 1, 72840 Puebla, Mexico
\and
            Department of Physics and Astronomy, University of California, Davis, CA 95616, USA
\and
            University of Massachusetts, Amherst, MA 01003, USA
\and
            Rochester Institute of Technology, Rochester, NY 14623, USA
\and
            Ben-Gurion University of the Negev, P.O. Box 653, Be’er-Sheva 8410501, Israel
          }

\abstract{Galaxy cluster mergers are representative of a wide range of physics, making them an excellent probe of the properties of dark matter and the ionized plasma of the intracluster medium. To date, most studies have focused on mergers occurring in the plane of the sky, where morphological features can be readily identified. To allow study of mergers with arbitrary orientation, we have assembled multi-probe data for the eight-cluster ICM-SHOX sample sensitive to both morphology and line of sight velocity. The first ICM-SHOX paper \cite{silich2023} provided an overview of our methodology applied to one member of the sample, \macscom in order to constrain its merger geometry. That work resulted in an exciting new discovery of a velocity space decoupling of its gas and dark matter distributions. In this work, we describe the availability and quality of multi-probe data for the full ICM-SHOX galaxy cluster sample. These datasets will form the observational basis of an upcoming full ICM-SHOX galaxy cluster sample analysis. }

\maketitle

\section{Introduction}
\label{intro}

Galaxy clusters represent the current stage of cosmological structure formation, and major mergers between clusters of similar mass ($R = M_{\text{primary}}/M_{\text{secondary}} \lesssim$ 10) are the primary hierarchical growth mechanism of these objects \cite{muldrew15, Molnar2016}. Such cluster mergers play a dual role as probes of astrophysics at extreme scales and as diagnostics of underlying cosmological parameterizations (e.g., \cite{lacey1993, Fakhouri2010, Thompson2012}). The first ICM-SHOX paper \cite{silich2023} introduced ICM-SHOX (Improved Constraints on Mergers with Sunyaev-Zel’dovich (SZ), Hydrodynamical simulations, Optical, and X-ray), which comprises a sample of massive galaxy clusters and an analysis pipeline that jointly infers a set of characteristic merger parameters by employing a framework that directly compares a novel combination of multi-probe cluster observables to analogous mock observables derived from a suite of hydrodynamical simulations. As described in \cite{silich2023}, the plane-of-sky (POS) morphology and line-of-sight (LOS) velocity structure of the gas (intracluster medium; ICM) and dark matter (DM) distributions of each cluster in the ICM-SHOX sample are characterized by the following observables:

\begin{itemize}     \item \textbf{POS DM morphology:} mass reconstructions derived from \textit{Hubble Space Telescope} (\textit{HST}) gravitational lensing (GL) measurements map the spatial distribution of DM (e.g., \cite{zitrin2011, zitrin2015})
    \item \textbf{POS ICM morphology:} \textit{Chandra} X-ray observations and thermal SZ (tSZ) effect data from \textit{Planck} and the ground-based \textit{AzTEC} and \textit{Bolocam} instruments \cite{sayers2019} probe the thermodynamics and morphology of the ICM. 
    \item \textbf{LOS DM velocity:} spectroscopic redshifts from cluster-member galaxies obtained with \textit{Keck/DEIMOS} and \textit{Keck/LRIS} trace the DM velocity structure (e.g., \cite{Ma2009, Owers2011, Boschin2013}). 
    \item \textbf{LOS ICM velocity:} kinematic SZ (kSZ) effect data from \textit{Planck}, \textit{AzTEC} and \textit{Bolocam} \cite{sayers2019} trace the ICM velocity structure. 
\end{itemize}

The ICM-SHOX pipeline then generates a set of analogous mock observables tailored to each of these probes from a suite of idealized hydrodynamical binary galaxy cluster merger simulations. These simulations, characterized by a vast array of initial parameters, are run with the \texttt{GAMER-2} code \cite{Schive2018}. Finally, the framework jointly infers a set of likely characteristic merger parameters for each cluster using a frequentist statistical analysis. Applying the ICM-SHOX pipeline to the multi-probe data available for \macscom \cite{silich2023} constrained the overall geometry of the merger, providing quantitative estimates of the component masses and gas profiles, the merger viewing angle, epoch, impact parameter, and initial relative velocity of the two subclusters. They further discovered a velocity space decoupling of the gas (ICM) and DM distributions in \macscom which the simulations predict for particular merger epochs, geometries, and viewing angles. 

The analysis of \cite{silich2023} demonstrated the ability of the ICM-SHOX multi-probe data and analysis pipeline to determine cluster merger parameters, as well as an example of exciting new astrophysical phenomena that can be uncovered through these novel methods. In this complementary work, we describe the available data and relevant quality metrics for all galaxy clusters in the ICM-SHOX sample. 

\section{The ICM-SHOX galaxy cluster sample}
\label{sec:sample}

The ICM-SHOX sample comprises eight massive (M$_{500} \gtrsim 10^{15}$ M$_{\odot}$), intermediate-redshift ($0.25 \lesssim z \lesssim 0.6$) galaxy clusters originally identified in \cite{sayers2019}. The systems were primarly selected as major mergers (e.g., MACS J0717.5+3745, a very massive, complex system with four merging subclusters; \cite{Ma2009}). The ICM-SHOX sample spans a broad range of presumed merger inclination angles and dynamical states, with four cluster mergers likely occurring primarily along the LOS (Abell 0697, \macscom MACS J0717.5+3745, and MACS J2129.4$-$0741), three mergers likely occurring near the POS (MACS J0025.4$-$1222, MACS J0454.1$-$0300, and RX J1347.5$-$1145), and one dynamically relaxed object (Abell 1835) \cite{sayers2019}. Population statistics of characteristic merger parameters in these systems will allow us to test for deviations from $\Lambda$CDM in the most extreme regime of cosmological structure formation and growth. In Table \ref{tab:data_table}, we detail the relevant physical properties of each cluster and corresponding quality metrics for the available multi-probe data. 

\begin{landscape}
\begin{table} [ht]
\centering
\caption{ICM-SHOX galaxy cluster sample details: physical properties and data quality metrics}
\label{tab:data_table}
\begin{tabular}{lccccccc}
\hline\hline
    \underline{Cluster} & \underline{$z$} & \underline{M$_{500}$} & \underline{$X_S$} & \underline{$z_{\text{spec}}$} & \underline{v$_\text{RMS}$} & \underline{$\tau_\text{RMS}$} & \underline{GL RMS$_{}$} \vspace{1mm}\\ 
    &  & [$10^{14}$ M$_{\odot}$] & [$\times 10^3$] & [$N_{\text{gal}}$]  & [$\times 10^3$ km s$^{-1}$] & [$\times 10^{-3}$] & [$''$] \vspace{1mm} \\ \hline
    Abell 0697 & 0.28  & 17.1 & $\phantom{1}$14.0  & 104$^{a}$ \cite{Girardi2006} & 1.4  & 2.1 & $\phantom{1}$0.82 \cite{cibirka2018} \\
    Abell 1835 & 0.25  & 12.3 & 342.0  & $\phantom{1}$60$^{a}$ \cite{Rines2013} & 0.9  & 2.7 & 1.4 \cite{Morandi2012} \\
    \macs & 0.55  & 16.5 & $\phantom{1}$18.4  & $\phantom{1111}$156 \cite{obsz_dressler, obsz_ellingson, obsz_crawford, silich2023} &  0.8 & 1.5 & $\phantom{1}$0.86 \cite{furtak2022} \\
    MACS J0025.4$-$1222  & 0.58  & $\phantom{1}$7.6  & $\phantom{1}$16.6  &  KOA$^{b}$ &  1.2 &  1.6 & $\phantom{1}$0.57 \cite{cibirka2018} \\
    MACS J0454.1$-$0300  & 0.54  & 11.5 & $\phantom{1}$21.1  & 149 \cite{Moran2007}$\phantom{1}$ &  1.1 &  1.6 & $\phantom{1}$0.60 \cite{jauzac2021}  \\
    MACS J0717.5+3745 & 0.55  & 24.9 & $\phantom{1}$63.4  & 237 \cite{Ma2008}$\phantom{1}$ & 0.8  & 1.2 & 3.18 \cite{zitrin2015} \\
    MACS J2129.4$-$0741 & 0.59  & 10.6 & $\phantom{111}$2.9$^{c}$  & 131 \cite{Ebeling2007}$\phantom{1}$ & 1.2  & 1.2 & 2.42 \cite{zitrin2015} \\
    RX J1347.5$-$1145  & 0.45  & 21.7  & 214.2  & $\phantom{11}$104 \cite{foex2017, Cohen2002}  & 0.8  &  1.4 & 2.61 \cite{zitrin2015} \\ \hline\hline \\ \vspace{-4mm}
\end{tabular} \\ 
\raggedright 
\textbf{Notes.} \textit{Column 1:} Galaxy cluster name. \textit{Column 2:} Cluster redshift. \textit{Column 3:} X-ray-derived cluster mass M$_{500}$. \textit{Column 4:} Number of cluster source counts in existing \textit{Chandra} data. \textit{Column 5:} Number of cluster-member spectroscopic redshifts used to generate corresponding maps of cluster-member galaxy velocities. \textit{Columns 6--7:} median RMS uncertainty of the central four pixels in the SZ data products (ICM v$_{\text{pec}}$ and $\tau$ maps). \textit{Column 8:} GL image-plane reproduction RMS. Data from columns 2, 3, 4, 6, and 7 came from \cite{sayers2019}. References for other columns are individually noted. Upon completion of the planned \textit{Keck} and \textit{Chandra} observations, the effective uncertainties on the multi-probe observables will generally vary by a factor of $\lesssim 2$ over the full sample, except for the gravitational lensing models and the much deeper \textit{Chandra} observations of Abell 1835 and RX J1347.5$-$1145. \\ \vspace{3mm}
$^{a}$Further data collection is in progress as part of our ongoing \textit{Keck} campaign (Proposal ID: C334; PI: S. Golwala). \\ \vspace{1mm}
$^{b}$Reduction of \textit{Keck Observatory Archive} (KOA) data is in progress now. \\ \vspace{1mm}
$^{c}$170 ks of \textit{Chandra} observations (Proposal ID: 24800077; PI: J. Sayers) are currently in progress, with which we expect to increase the MACS J2129.4 source counts to $\simeq15 \times 10^3$.
\end{table}
\end{landscape}

In the following figures (\ref{fig:vgals}--\ref{fig:lensing}), we provide examples of observables derived from the data highlighted in Table \ref{tab:data_table} for select clusters in ICM-SHOX. 

\begin{figure}[ht]
\centering
\includegraphics[width=1\linewidth]{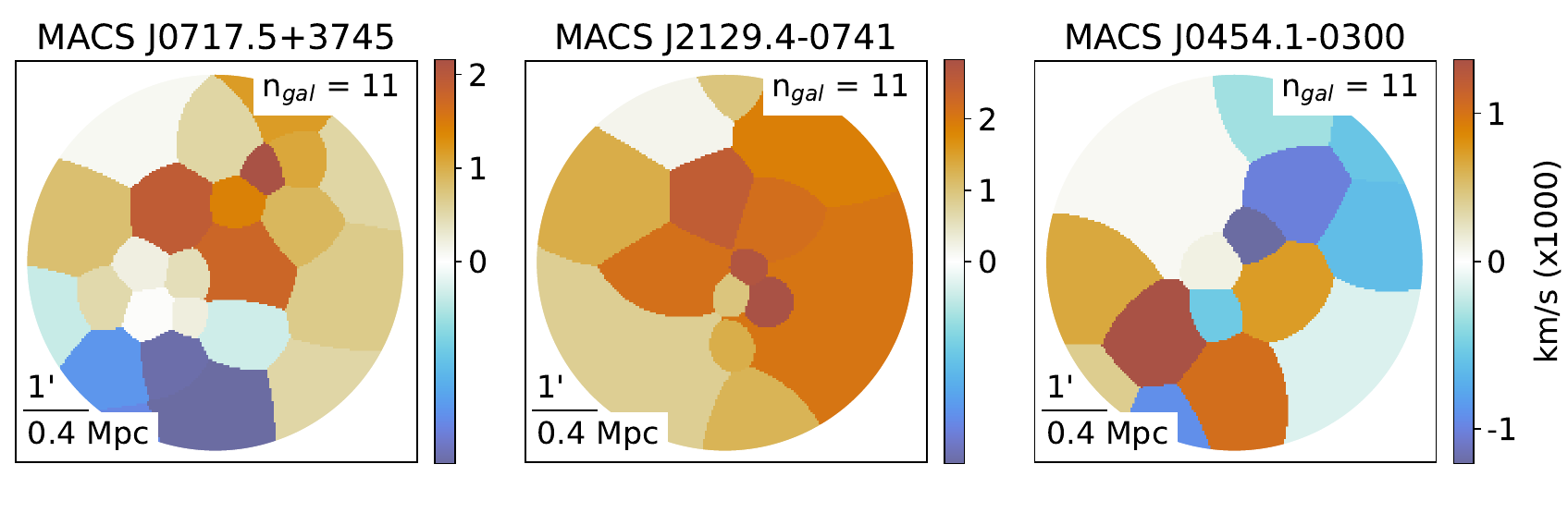}
\caption{Example maps of cluster-member galaxy velocities ($v_{\text{gal}}$), which are assumed to trace the LOS DM velocity structure in each cluster. The $v_{\text{gal}}$ map generation procedures are detailed in \cite{silich2023}.}
\label{fig:vgals}  
\end{figure}

\begin{figure}[ht]
\centering
\includegraphics[width=1\linewidth]{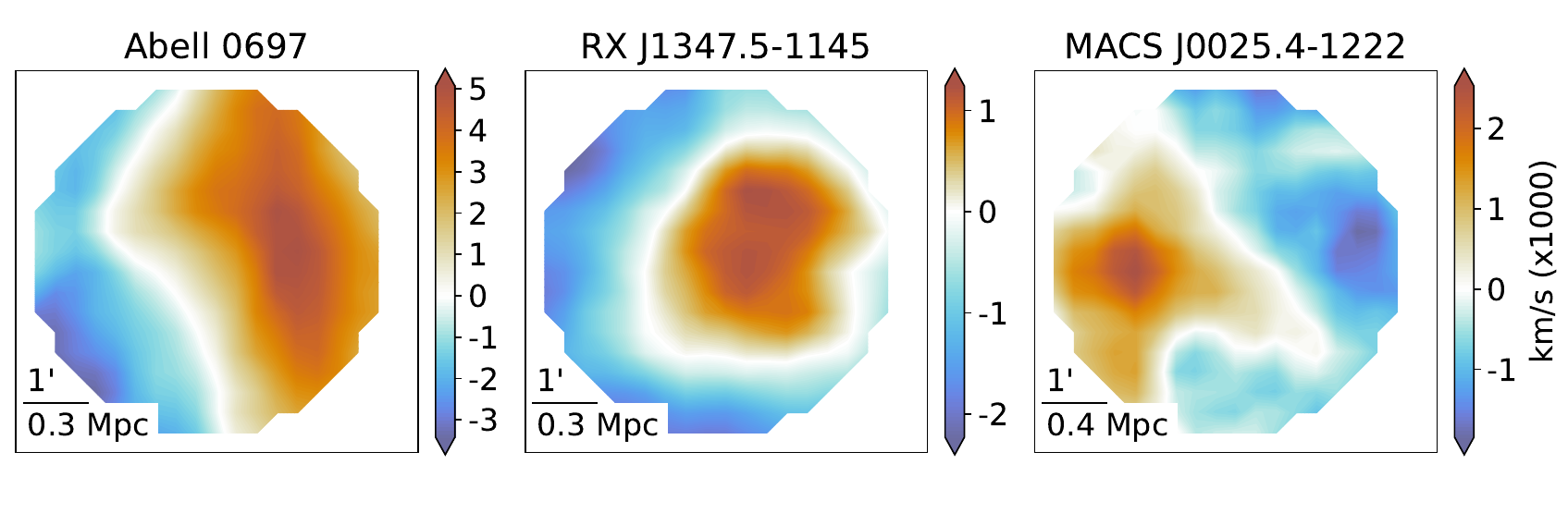}
\caption{Example maps of the ICM velocity structure (ICM $v_{\text{pec}}$) projected along the LOS. The ICM $v_{\text{pec}}$ maps for each cluster were originally generated in \cite{sayers2019} using kSZ imaging data.}
\label{fig:vpec}  
\end{figure}

\begin{figure}[h!]
\centering
\includegraphics[width=0.99\linewidth]{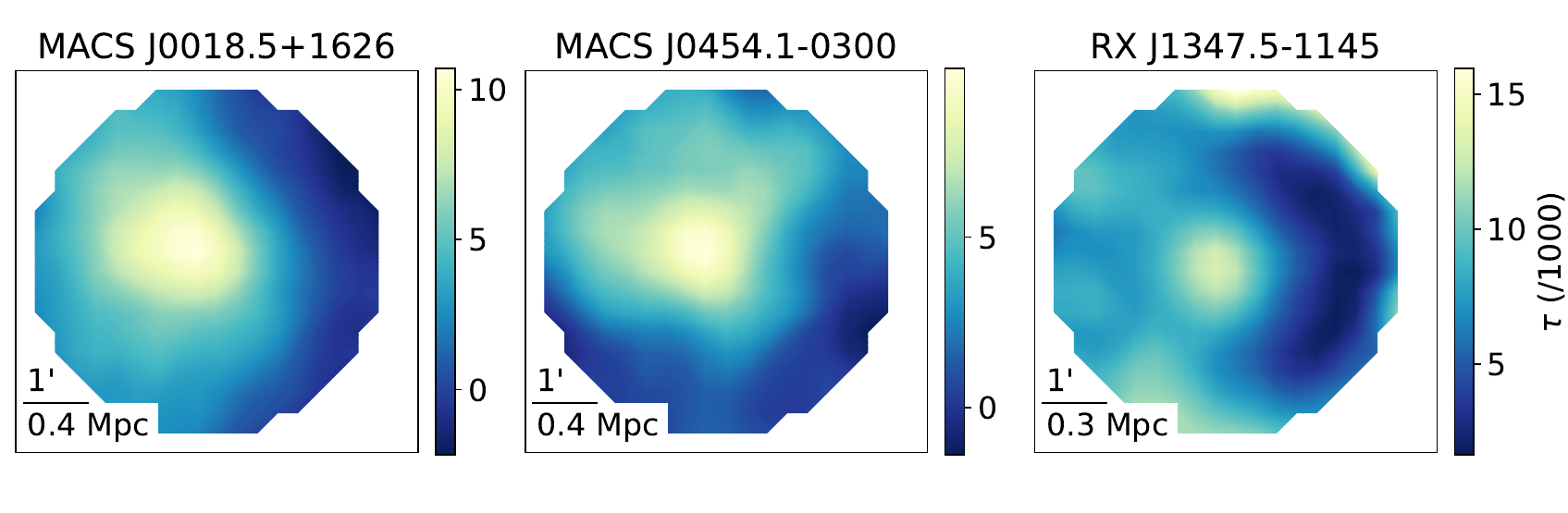}
\caption{Example maps of the electron optical depth ($\tau$) projected along the LOS. The $\tau$ maps for each cluster were originally generated in \cite{sayers2019} using kSZ imaging data.}
\label{fig:tau}  
\end{figure}

\begin{figure}[ht]
\centering
\includegraphics[width=1.03\linewidth]{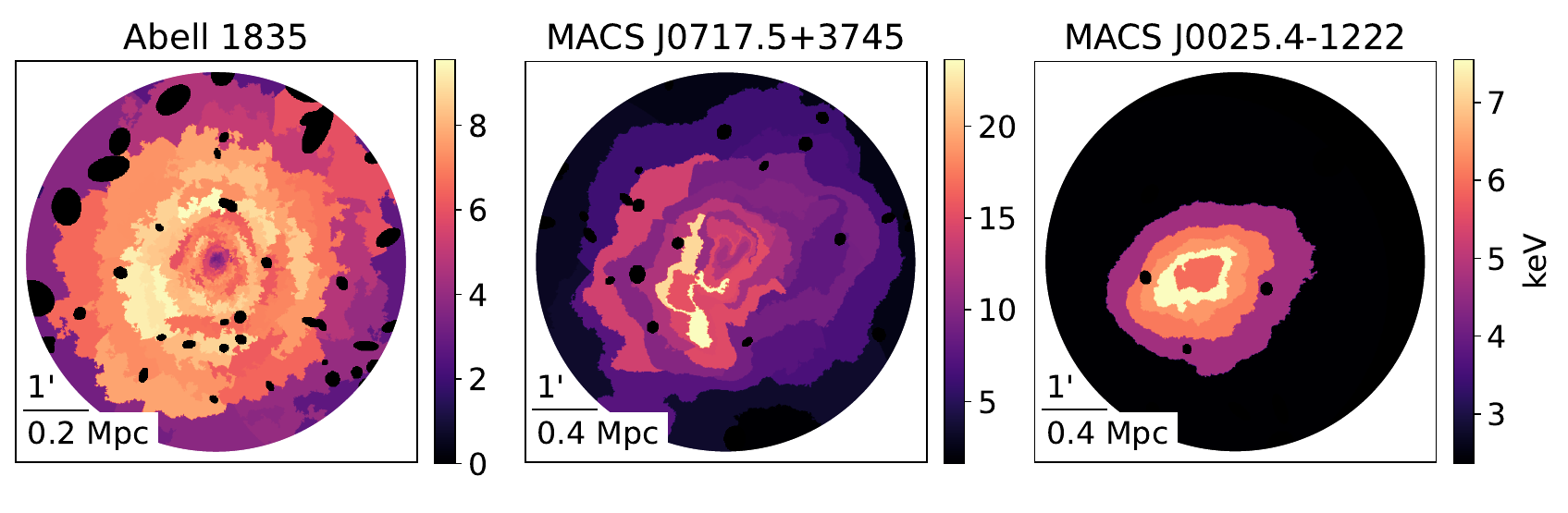}
\caption{Example maps of the X-ray derived temperature (kT). The kT maps for each cluster were originally generated in \cite{sayers2019} using \textit{Chandra} data.}
\label{fig:kT}  
\end{figure}

\begin{figure}[ht]
\centering
\includegraphics[width=1.03\linewidth]{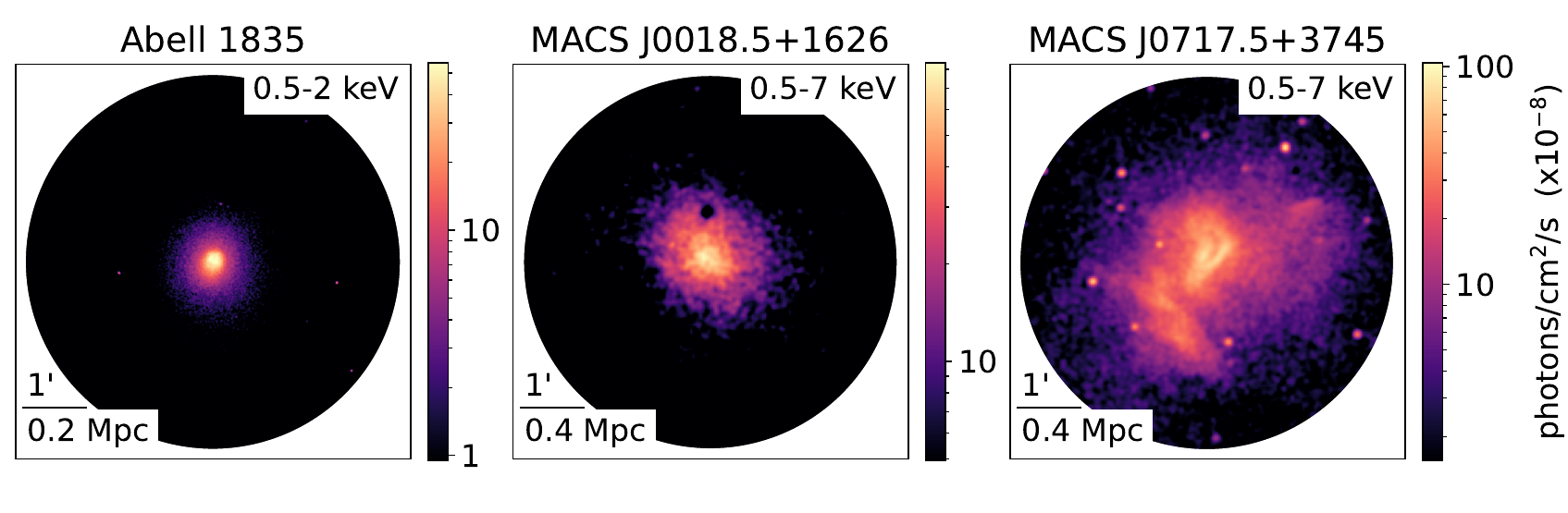}
\caption{Example maps of the X-ray surface brightness (XSB). The XSB maps for each cluster were originally generated in \cite{sayers2019} using \textit{Chandra} data, and the \macs map was updated in \cite{silich2023}.}
\label{fig:xsb}  
\end{figure}

\begin{figure}[h!]
\centering
\includegraphics[width=1.03\linewidth]{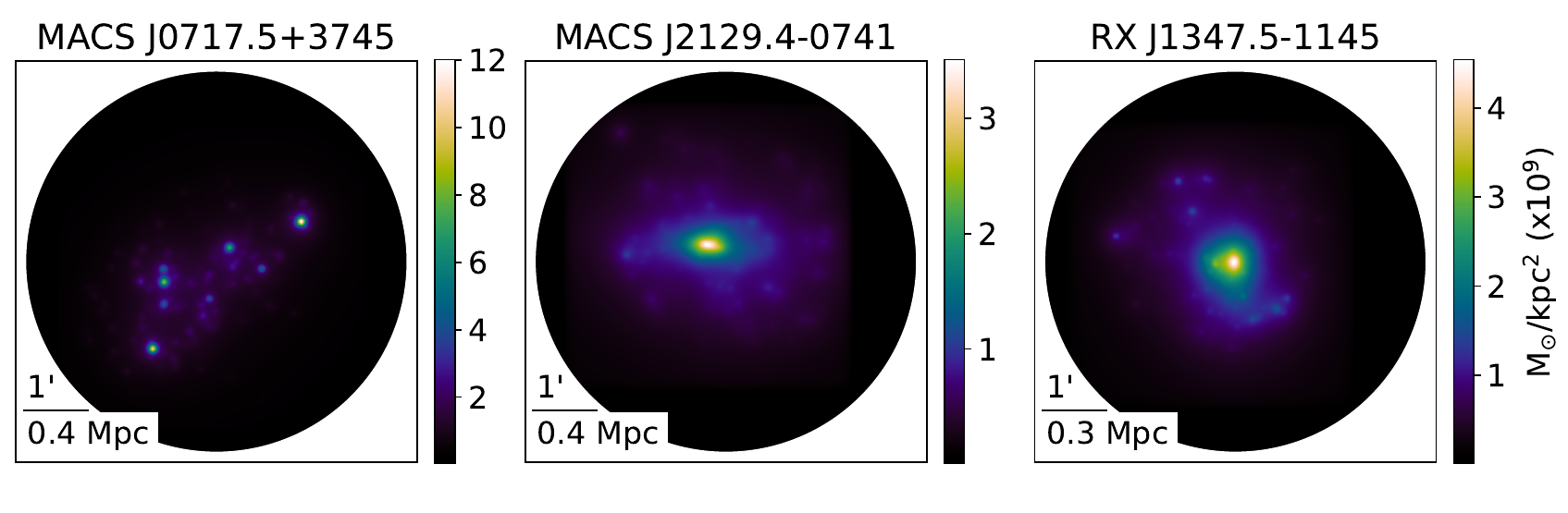}
\caption{Example maps of the total cluster mass ($\Sigma$) projected along the LOS. These $\Sigma$ maps were originally generated in \cite{zitrin2015} using multi-band \textit{HST} imaging data. }
\label{fig:lensing}  
\end{figure}

Some of the data presented herein were obtained at the W. M. Keck Observatory, which is operated as a scientific partnership among the California Institute of Technology (Caltech), the University of California, and the National Aeronautics and Space Administration (NASA). The Observatory was made possible by the generous financial support of the W. M. Keck Foundation. This research has also made use of the Keck Observatory Archive (KOA), which is operated by the W. M. Keck Observatory and the NASA Exoplanet Science Institute (NExScI), under contract with NASA. The authors wish to recognize and acknowledge the very significant cultural role and reverence that the summit of Maunakea has always had within the indigenous Hawaiian community.  We are most fortunate to have the opportunity to conduct observations from this mountain. The X-ray observations were obtained from the Chandra Data Archive. We acknowledge Harald Ebeling for providing redshifts for MACS J2129.4$-$0741 and MACS J0717.5+3745. 

We acknowledge financial support as follows: EMS: National Science Foundation Graduate Research Fellowship (NSF GRFP) under Grant No. DGE‐1745301, the Wallace L. W. Sargent Graduate Fellowship at Caltech, and NSF/AST-2206082; JS: NSF/AST-2206082; JAZ: the Chandra X-ray Center, which is operated by the Smithsonian Astrophysical Observatory for and on behalf of NASA under contract NAS8-03060; EB: NSF/AST-2206083; AZ: Grant No. 2020750 from the United States-Israel Binational Science Foundation (BSF) and Grant No. 2109066 from the United States NSF, and by the Ministry of Science \& Technology, Israel; TM: the AtLAST project, which has received funding from the European Union’s Horizon 2020 research and innovation program under grant agreement No 951815; AM: Consejo Nacional de Humanidades Ciencias y Technolog\'ias (CONAHCYT) project A1-S-45680.

\end{document}